\documentclass[pre,preprint, groupedaddress, showpacs]{revtex4}

\usepackage{graphicx}
\begin{document}


\title{Interaction of multi-armed spirals in bistable media}


\author{Ya-feng He$^{1}$}
\email[Email:]{heyf@hbu.edu.cn}
\author{Bao-quan Ai$^{2}$}
\author{Fu-cheng Liu$^{1}$}


\affiliation{$^1$Hebei Key Laboratory of Optic-electronic Information Materials, College of Physics Science and Technology, Hebei University, Baoding 071002, China\\
$^2$Laboratory of Quantum Information Technology, ICMP and SPTE, South China Normal University, Guangzhou 510006, China.}


\date{\today}
\begin{abstract}
\indent We study the interaction of both dense and sparse multi-armed spirals in bistable media modeled by equations of FitzHugh-Nagumo type. Dense 1-armed spiral is characterized by its fixed tip. For dense multi-armed spirals, when the initial distance between tips is less than a critical value, the arms collide, connect and disconnect continuously as the spirals rotate. The continuous reconstruction between the front and the back drives the tips to corotate along a rough circle and to meander zigzaggedly. The rotation frequency of tip, the frequency of zigzagged displacement, the frequency of spiral, the oscillation frequency of media, and the number of arms satisfy certain relations as long as the control parameters of the model are fixed. When the initial distance between tips is larger than the critical value, the behaviors of individual arms within either dense or sparse multi-armed spirals are identical to that of corresponding 1-armed spirals.
\end{abstract}



\maketitle
\section {Introduction}
\indent Spiral waves have been observed in a variety of physical, chemical, and biological systems \cite{Cross, Kepper, Epstein, Petrov, Nattel, Gurevich, Chen, Dong}. Although the intrinsical origins of the spirals in these systems are very different, the dynamics of the spirals exhibit remarkable similarities, such as the meandering \cite{Ouyang,Barkley}. From the point of view of dynamics, the spirals can be classified into three types: excitable, oscillatory, and bistable spirals. The excitable spirals originate from the excitability of media, which have been studied widely in the Belousov-Zhabotinsky (BZ) reaction \cite{Kupitz}, the cardiac tissue \cite{Nattel}, and the catalytic reactions on platinum surface  \cite{Bar}. The oscillatory spirals originate from the Hopf bifurcation and are formed by the coupling of phase in spatial. They have been investigated extensively in the BZ reaction \cite{Ouyang}, the nonlinear optical system \cite{Tlidi}, the complex Ginzburg-Landau equation \cite{Tlidi, Xie}, and the FitzHugh-Nagumo model \cite{Mau}. The bistable spirals consist of two Bloch fronts. Their dynamics are entirely determined by the Bloch interfaces \cite{Hagberg, Hagberg1}. The bistable spirals have been observed in the ferrocyanide-iodate-sulfite reaction \cite{Kepper, Li}, and the FitzHugh-Nagumo model \cite{He2, Hagberg2}. From the point of view of profile, the spirals show dense or sparse configuration \cite{Gottwald, Krinsky, He}. In excitable media, dense and sparse spirals result from different excitabilities of media. In bistable media, dense and sparse spirals can transit each other via a subcritical bifurcation. So far, the dynamics of 1-armed spirals have been well understood.

\indent Recently major progress on the spirals has involved the interaction of multi-armed spirals \cite{Ginn, Bursac, Vasiev, Zemlin1, Zemlin2, Zaritski}. This is inspired especially by the experimental observation of multi-armed spirals in a two-dimensional cardiac substrate \cite{Bursac}. The behaviors of persistent multi-armed spirals are different from that of 1-armed spirals. Multi-armed spirals can be formed due to attraction of several 1-armed spirals if their tips are less than one wavelength apart \cite{Vasiev}. In an asymmetric bound state of spiral pairs, the faster spiral becomes the master one, and the other (slave spiral) can rotate around it \cite{Zemlin2}. Although the study on the interaction of multi-armed spirals has rose in the past years, it has mainly focused on the sparse spirals in excitable media. The interaction of multi-armed spirals in bistable media, for both dense and sparse configurations, is still an open question. In this work, we first give the evolution of both dense and sparse 1-armed spirals. Then, we study the interaction of dense multi-armed spirals in detail, including the repulsive and attractive interactions, the corotation and zigzagged meandering of tips, and the relations between several frequencies. Finally, we discuss the interaction of sparse multi-armed spirals.

\section{BISTABLE MEDIA MODEL}

\indent This work is based on the FitzHugh-Nagumo model,
\begin{eqnarray}
  u_t &=& au -u^3 - v + \nabla^2 u,\\
  v_t &=& \varepsilon(u-v)+\delta \nabla^2 v,
\end{eqnarray}

here variables $u$ and $v$ represent the concentrations of the activator and inhibitor, respectively, and $\delta$ denotes the ratio of their diffusion coefficients. The small value $\varepsilon$ characterizes the time scales of the two variables. In this paper the parameter $a$ is chosen such that the system is bistable. The two stationary and uniform stable states are indicated by up state ($u_{+}$,$v_{+}$) and down state ($u_{-}$,$v_{-}$), respectively, and they are symmetric, $(u_{+},v_{+})$=$-(u_{-},v_{-})$. An interface connects the two stable states smoothly. On decreasing $\varepsilon$ the system follows nonequilibrium Ising-Bloch bifurcation that leads to the formation of a couple of Bloch interfaces \cite{He2, Hagberg2}. In the following, we define a front which jumps from down to up state, and a back which falls from up to down state. Here the front and the back correspond to the two Bloch interfaces with opposite velocities. So the image of bistable spiral is clear: a couple of Bloch interfaces (front and back) propagating with opposite velocities enclose a spiral arm, and the front meets the back at the spiral tip \cite{Hagberg, Hagberg1}. In the following the spiral tip is determined by the intersection of the single contour of $u$ and $v$, at where $u$$=$$v$$=$$0$. The bistable system can exhibit dense and sparse spirals. In order to differentiate them, we define an order parameter $\alpha$$=$$|\frac{\lambda_{+}-\lambda_{-}}{\lambda_{+}+\lambda_{-}}|$, here $\lambda_{+}$ and $\lambda_{-}$ represent the average widths of up state and down state, respectively. In dense spiral case, $\alpha$$=$$0$ which means that the up and down states have identical widths [as Fig. 1(a) shown]. In sparse spiral case, if $\lambda_{+}$$<$$\lambda_{-}$ we call it Positive Phase Sparse Spiral [PPSS, Fig. 1(b)]. Otherwise, if $\lambda_{+}$$>$$\lambda_{-}$ we call it Negative Phase Sparse Spiral [NPSS, Fig. 1(c)].

\indent Spiral waves in bistable media are sensitive to the initial condition and the boundary condition. Multi-armed spirals can be generated by using the initial conditions with different number of triangles as shown in Figs. 1(d) and 1(h). The vertices of the triangles locate on an initial circle with different radius $r$ and serve as the initial tips of generated spirals. They are separated equally by different angles in order to generate the spirals with different number of arms. For 2-armed spiral the two vertices are separated by $\pi$ as shown in Fig. 1(h). For 3-armed spiral they are separated by $2$$\pi$/$3$, and so on. Figure 1(d) gives the initial condition for 1-armed spiral. It should be mentioned that the numerical results will remain unchanged for slight deviation of the given initial conditions. The boundary conditions are taken to be no-flux. A generalized Peaceman-Rachford ADI scheme is used to integrate the above model. Unless otherwise noted, our simulations are under the parameter sets: $a$$=$$2.0$, $\delta$$=$$0.05$, space step $dx$$=$$dy$$=$$0.4$ space units, time step $dt$$=$$0.04$ time units and domain size $L$$\times$$L$$=$$160$$\times$$160$ space units.

\section{Simulation results and discussion}

\subsection{1-armed spiral}

\begin{figure}[htbp]
  \begin{center}\includegraphics[width=8cm,height=4cm]{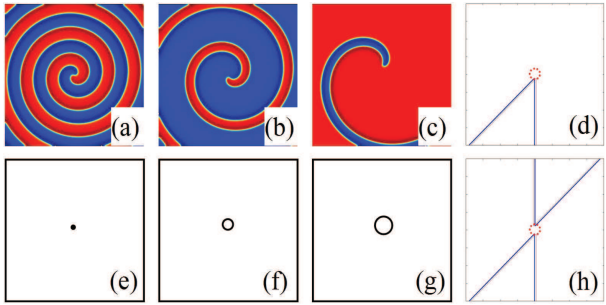}
  \caption{(Color online) Single-armed spiral in bistable media. (a) dense spiral, $\varepsilon$$=$$0.2$; (b) Positive Phase Sparse Spiral (PPSS), $\varepsilon$$=$$0.36$; (c) Negative Phase Sparse Spiral (NPSS), $\varepsilon$$=$$0.4$; (e)-(g) corresponding tip paths. (d) and (h) indicate the initial conditions for generating 1-armed and 2-armed spirals. The thick (thin) lines in (d) and (h) represent the contour lines $u$=$0$ ($v$=$0$), and the dashed circle indicates the initial circle on which the initial tips located. Domain size:$80$$\times$$80$ space units.}
  \label{1}
  \end{center}
\end{figure}

\indent We first show the evolution of the 1-armed spiral with increasing $\varepsilon$ as shown in Figs. 1(a)-(c). In order to obtain 1-armed spiral we use the initial condition as indicated in Fig. 1(d). When $\varepsilon$$<$$0.33$ the observed spiral is dense as shown in Fig. 1(a), in which the order parameter $\alpha$$=$$0$ and does not depend on epsilon according to its definition. The tip of the dense spiral keeps a fixed point as shown in Fig. 1(e). When $\varepsilon$$\geq$$0.33$ the observed spiral is sparse as shown in Fig. 1(b) and 1(c). The order parameter $0$$<$$\alpha$$<$$1$ and increases with epsilon. The tip of the sparse spiral follows a circle. The radius of this circle increases with $\varepsilon$. The trajectories of the tips in this case are shown in Figs. 1(f) and 1(g), respectively. The transition between dense spiral and sparse spiral proves to be a subcritical bifurcation \cite{He}.

\subsection{Multi-armed dense spirals}

\begin{figure}[htbp]
  \begin{center}\includegraphics[width=7cm,height=7cm]{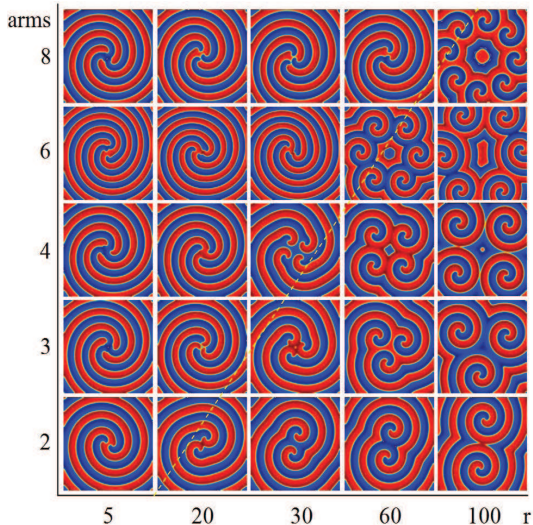}
  \caption{(Color online) Multi-armed dense spirals with different number of arms generated from different initial circles. Dashed line divides these spirals into two groups.}
  \label{2}
  \end{center}
\end{figure}

\indent Our emphasis is on the interaction between the arms within a multi-armed dense spiral ($\varepsilon$$<$$0.33$). The tips of multi-armed dense spirals can corotate along a rough circle or fix at their individual points, which depends on their initial separations. Figure 2 shows the multi-armed dense spirals with different number of arms generated from different initial circles when $\varepsilon$$=$$0.2$. The horizontal axis represents the radius of the initial circle as indicated in Fig. 1(h), which shows the initial separation of the tips. The vertical axis represents the number of the initial tips spaced equally on the initial circle. It can be seen that when the distance between initiated tips is larger enough, the arms will rotate independently as shown on the right of the dashed line in Fig. 2. In this case, the number of the arms within the final spiral is equal to that of the initial arms. When the distance between initiated tips is small, the strong interaction between the arms leads to the instability of the multi-armed spiral. Some of the arms become unstable and run outside of the domain in the stationary state. The rest of the arms then begin to corotate with each other as is indicated on the left of the dashed line in Fig. 2. The number of the arms within the final spiral is less than that of the initial arms. So, the dashed line in Fig. 2 divides these spirals into two groups. It means that there exists a critical radius of the initial circle below which the arms begin to interact strongly with each other. Figure 3 gives the dependence of this critical radius $r_{c}$ on the parameter $\varepsilon$. It is clear that the critical radius increases with $\varepsilon$. Moreover, for the same $\varepsilon$, the more the initial arms are, the larger the critical radius is. This is because that, for the spiral with more arms, it becomes easier to interact.

\begin{figure}[htbp]
  \begin{center}\includegraphics[width=7cm,height=5cm]{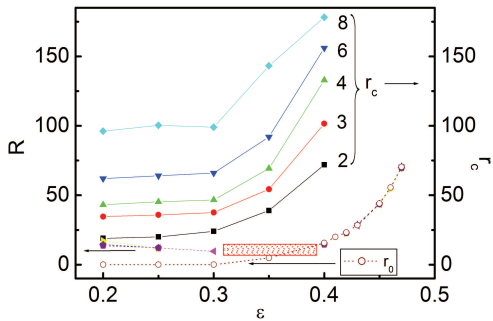}
  \caption{(Color online) Dependence of the critical radius $r_{c}$, the radius $R$ of the stationary circle on the parameter $\varepsilon$. Solid lines (dashed lines) represent the radius $r_{c}$ ($R$). Circle dotted line $r_{0}$ represents the radius of the tip trajectory of 1-armed spiral. The patterned rectangle region near the bifurcation point $\varepsilon$$=$$0.33$ shows very complex interaction between the arms. The digits in figure indicate the numbers of the arms. When the radius of the initial circle is less than the critical radius, i.e., $r$$<$$r_{c}$, strong interaction between arms results in the corotation of tips along one zigzagged circle with radius $R$. When $r$$>$$r_{c}$ the arms of multi-armed spirals will rotate independently like 1-armed spirals.}
  \label{3}
  \end{center}
\end{figure}

\begin{figure}[htbp]
  \begin{center}\includegraphics[width=7cm,height=6cm]{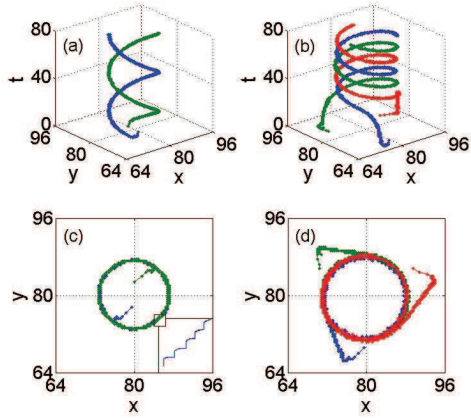}
  \caption{(Color online) Rough profiles of the tip trajectories for 2-armed and 3-armed dense spirals. (a) and (b) plot the trajectories in $x$, $y$, and $t$ space. (c) and (d) plot them in $x$, $y$ space. The initial separation of the tips for 2-armed spiral (3-armed spiral) is $r$$=$$5$ ($r$$=$$26$). Smaller (larger) separation between the initial tips exhibits repulsive (attractive) interaction. Inset in (c) represents the zigzagged meandering of one tip.}
  \label{4}
  \end{center}
\end{figure}

\indent The strong interaction of multi-armed dense spirals can be studied by following the tip trajectories. Fig. 1(a) and Fig. 2 show dense 1-armed spiral and multi-armed spiral at $\varepsilon$$=$$0.2$, respectively. For 1-armed dense spiral [Fig. 1(a)] the tip is stable as indicated by the fixed point in Fig. 1(e). For multi-armed dense spirals located on the left of the dashed line in Fig. 2, the tips follow a rough circle and corotate with each other due to their interaction. Figure 4 gives two examples which show the rough profiles of the tip trajectories for 2-armed and 3-armed dense spirals, respectively. Figs. 4(a) and 4(c) show the space-time plots of the two tips within 2-armed spiral, in which the two tips are initiated on an initial circle with smaller radius $r$$=$$5$. It is clear that the two tips repel each other at first, and then corotate with each other along a circular stationary trajectory with radius $R_{2}$$=$$14$. However, if the tips are initiated on an initial circle with larger radius ($r$$>$$R_{2}$, but $r$$<$$r_{c}$), they firstly attract each other and then corotate along the circular stationary trajectory with the same radius $R_{2}$$=$$14$. Figs. 4(b) and 4(d) show the attractive interaction for 3-armed spiral, in which the three tips are initiated on an initial circle with larger radius $r$$=$$26$. The three tips finally corotate along a circle with radius of $R_{3}$$=$$17.2$. Similarly, if $r$$<$$R_{3}$, the tips exhibit repulsive interaction, and also follow the circular stationary trajectory with radius $R_{3}$$=$$17.2$. Here, we should note that the radius $R_{3}$$>$$R_{2}$. This is because that the core region of 3-armed spiral is larger than that of 2-armed spiral due to their repulsive interaction.

\begin{figure}[htbp]
  \begin{center}\includegraphics[width=7cm,height=5cm]{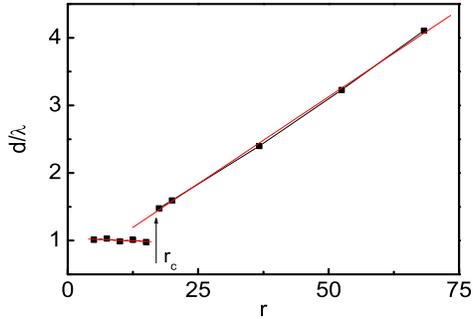}
  \caption{(Color online) Dependence of $d$$/$$\lambda$ on $r$ in 2-armed spiral. Here, $d$ represents the average distance between the two tips in final state. $\lambda$ is the wavelength of dense spiral. $r_{c}$ indicates the critical radius of initial circle. Straight lines show the linear fitting. When $r$$<$$r_{c}$ the strong interaction between the arms results in the tip corotation along a circular stationary trajectory. In this case, $d$$=$$\lambda$, which illustrates that the average distance between tips equals to the wavelength of dense spiral.}
  \label{5}
  \end{center}
\end{figure}

\indent We have conducted more numerical simulations for various number of initial arms and obtained the dependence of the radius $R$ of the stationary circle on parameter $\varepsilon$ as indicated by the dashed lines in Fig. 3. From Fig. 2 and Fig. 3 we can draw a conclusion that, as long as the radius $r$ of the initial circle is less than $r_{c}$ (solid lines in Fig. 3), i.e., $r$$<$$r_{c}$, the arms will interact strongly as shown on the left of the dashed line in Fig. 2. The tips can corotate with each other along a circular stationary trajectory with radius of $R$ (dashed lines in Fig. 3). The average distance $d$ between tips in the final state equals to the wavelength $\lambda$ of dense spiral as shown in Fig. 5. Otherwise, when $r$$>$$r_{c}$, the arms will rotate independently like 1-armed dense spiral as shown on the right of the dashed line in Fig. 2. The tips are fixed and separated well. Therefore, the multi-armed spirals on the right of the dashed line in Fig. 2 can be seen as coexistence of several 1-armed dense spirals. The average distance $d$ between tips in the final state increases linearly with the radius $r$ of the initial circle as shown in Fig. 5, which follows approximately that,

\begin{eqnarray}
  d/\lambda=0.535+0.052r.
\end{eqnarray}
The interaction condition in bistable media is different from that in excitable media. Vasiev $et$ $al$ have shown that multi-armed spirals in excitable media can corotate when their tips are less than one wavelength apart \cite{Vasiev}. However, in our case, only when $r$$<$$r_{c}$, multi-armed spirals can corotate, and meanwhile the tips will keep one wavelength apart as they corotate.

\begin{figure}[htbp]
  \begin{center}\includegraphics[width=7cm,height=10cm]{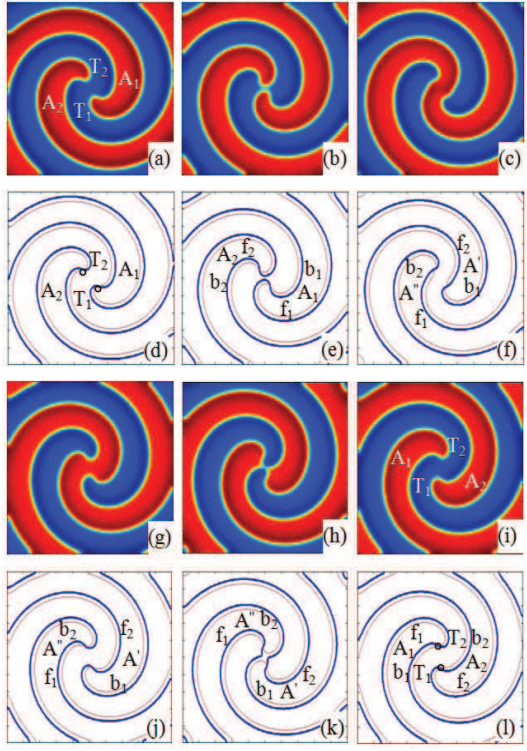}
  \caption{(Color online) Reconstruction of the front ($f_{i}$) and the back ($b_{i}$) in 2-armed dense spiral during one cycle. Thick (thin) lines in (d)-(f), (j)-(l) represent the corresponding contours of $u$ ($v$). Points of intersection between the thick and the thin lines indicate the two tips of 2-armed spiral, which is labeled by $T_{1}$ and $T_{2}$, respectively. The two arms (up state) in (d), (e) and (l) are labeled by $A_{1}$ and $A_{2}$, respectively. The two arms (down state) in (f), (j) and (k) are labeled by $A'$ and $A''$, respectively. Time in (a)-(c), (g)-(i) is $t$$=$$0$, $0.24$, $0.4$, $0.64$, $0.88$, $1.12$, respectively.}
  \label{6}
  \end{center}
\end{figure}

\indent The strong interaction of dense multi-armed spiral results in arm-switching with tips. Figure 6 shows the evolution of a corotating 2-armed spiral during one cycle. Firstly, we set the combination of the tips and the arms in Figs. 6(a) and 6(d) to $T_{1}$$-$$A_{1}$ and $T_{2}$$-$$A_{2}$. The two tips and the two arms separate well. During the development the two arms undergo collision [Figs. 6(b) and 6(e)], connection [Figs. 6(c), 6(f), 6(g), 6(j)], and disconnection [Figs. 6(h) and 6(k)]. After one cycle, the combination of the tips and the arms turns into $T_{1}$$-$$A_{2}$, $T_{2}$$-$$A_{1}$, and the arms switch to the other tip as shown in Figs. 6(i) and 6(l). This arm-switching will happen in the same way every one cycle.

\indent The interaction of multi-armed spirals and the mechanism of arm-switching can be studied by analyzing the interaction between the front and the back, because the arm of bistable spiral is enclosed by a couple of Bloch interfaces (a front and a back). In Fig. 6(e), the arm $A_{1}$ (up state) is enclosed by the front $f_{1}$ and the back $b_{1}$, and so does the other arm $A_{2}$. We should notice that when $\varepsilon$$=$$0.2$, for dense 1-armed spiral [Fig. 1(a)], the order parameter $\alpha$$=$$0$ ($\lambda_{+}$$=$$\lambda_{-}$). For dense 2-armed spiral [as Figs. 6(a) and 6(b)], it can be regarded as superposition of two sparse 1-armed spirals (arms $A_{1}$ and $A_{2}$, PPSS) with individual order parameter $\alpha_{1,2}$$=$$0.5$, and the global order parameter keeps $\alpha$$=$$0$. Because the parameter $\varepsilon$$=$$0.2$ the two sparse 1-armed spirals (arms $A_{1}$ and $A_{2}$, $\alpha_{1,2}$$=$$0.5$) tend to expand respectively to dense 1-armed spirals with $\alpha_{1,2}$$=$$0$. During their expansion the cores begin to collide and connect together [Figs. 6(b), 6(c), 6(g)], the front and the back will reconstruct. The back $b_{2}$ and the front $f_{1}$ enclose and form a new arm $A''$ (down state), the back $b_{1}$ and the front $f_{2}$ enclose and form the other new arm $A'$ (down state) as shown in Figs. 6(f) and 6(j). Now, the dense 2-armed spiral can also be regarded as superposition of two sparse 1-armed spirals (arms $A'$ and $A''$, NPSS). Similarly, the arms $A'$ and $A''$ will expand respectively as the rotation is going on, and the front and the back will reconstruct again. The front of one arm disconnects with the forward back of another arm, and reconnects with the backward back of itself [Figs. 6(i) and 6(l)]. Now, the combination of the tips and the arms in Figs. 6(i) and 6(l) is $T_{1}$$-$$A_{2}$, $T_{2}$$-$$A_{1}$, and the arms exchange their tips. So, the arm-switching observed in dense multi-arms spiral originates from the reconstruction of the front and the back. It is easy to deduce that undergoing one cycle again the arms will exchange their tips once more. The combination of the tips and the arms will become $T_{1}$$-$$A_{1}$ and $T_{2}$$-$$A_{2}$ like it was two cycles ago. In a word, the front and the back reconstruct twice in one cycle, which results in the arms-switching once. This mechanism of reconstruction happens also in other multi-armed dense spirals.

\begin{figure}[htbp]
  \begin{center}\includegraphics[width=7cm,height=6cm]{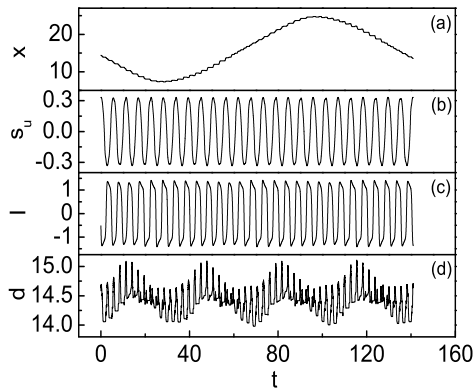}
  \caption{Time sequences of 3-armed dense spiral. (a):zigzagged displacement $x$ of one tip in $x$ direction. (b):parameter $s_{u}$. (c):oscillation intensity of one point far from the spiral center. (d):distance $d$ between two neighboring tips. The parameter $\varepsilon$$=$$0.25$, domain size $32$$\times$$32$ space units, space step $h$$=$$0.04$ space units, time step $dt$$=$$0.008$ time units.}
  \label{7}
  \end{center}
\end{figure}

\indent According to the eikonal equation, $c_{r}$$=$$c_{0}-D\kappa$, the curvature $\kappa$ of spiral affects the speed $c_{r}$ of the Bloch front and back. During the continuous reconstruction of front and back, the curvature of spiral near the tip changes remarkably as shown in Fig. 6(e) and 6(k). The changes in the curvatures of the front and the back result in the speed difference between them, which will drive the tip to zigzag along a rough circle (zigzagged meandering). The rotation direction of tip is determined by the chirality of spiral. Fig. 7(a) shows the time sequence of zigzagged displacement $x$ of one tip in $x$ direction for dense 3-armed spiral. It can be seen that the zigzagged displacement is tiny (about 0.36 space unit), which is far less than the average diameter of the circular stationary tip trajectory (about 14.28 space unit). In order to confirm if there exists petal meandering (like that in excitable media) in the present case, we use the domain size $28$$\times$$28$ space units and smaller space step $h$$=$$0.02$ space units. The numerical simulations with higher spatial resolution show only the zigzagged meandering of spiral. The zigzagged meandering of dense multi-armed spirals in bistable media is different from the petal meandering of spirals in excitable media. This is because the fact that the origins of meandering spirals in bistable and excitable media are different. The zigzagged meandering of multi-armed dense spiral in bistable media results from the interaction of the arms, i.e., the continuous reconstruction between the front and the back. The petal meandering of spiral in excitable media results from a secondary Hopf bifurcation \cite{Barkley}.

\indent In order to illustrate the periodical interaction between the arms, we define a parameter,

 \begin{eqnarray}
     s_{u}=\frac{1}{L^2}\sum_{i=1}^{L}\sum_{j=1}^{L} u_{i,j},
\end{eqnarray}

 where, $u_{i,j}$ represents the variable $u$ at the $(i,j)th$ grid point in two dimensional space. It can represent the frequency of dense multi-armed spiral. Fig. 7(b) shows the periodical change of this parameter $s_{u}$ for 3-armed spiral. When $s_{u}$$=$$0$, the up state and the down state are entirely symmetrical. We also measure the time sequence of one point far from the spiral center as shown in Fig. 7(c). Its frequency is far more than the rotation frequency of the tips [Fig. 7(a)]. The distance $d$ between two neighboring tips changes periodically as shown in Fig. 7(d). Obviously, there exist certain modulations of these time sequences in Fig. 7. In order to clarify these modulations, it is necessary to analyze their corresponding Fourier spectra.

\begin{figure}[htbp]
  \begin{center}\includegraphics[width=7cm,height=6cm]{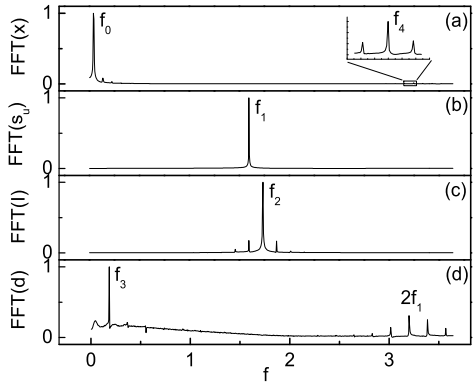}
  \caption{Fourier spectra of the corresponding time sequences of 3-armed dense spiral in Fig. 7.}
  \label{8}
  \end{center}
\end{figure}

\indent The zigzagged meandering of multi-armed dense spiral is characterized by several frequencies, the rotation frequency $f_{0}$ of tip, the frequency $f_{1}$ of spiral, the oscillation frequency $f_{2}$ of media, the frequency $f_{4}$ of zigzagged displacement, and the number $N_{arms}$ of arms. Because the tips meander clockwise, the oscillating frequency $f_{2}$ of one point far from the spiral center is larger than the frequency $f_{1}$ of spiral. The frequency $f_{2}$ is closely connected with the number $N_{arms}$ of arms. Obviously, the more arms one spiral has, the larger the frequency $f_{2}$ is. The rotation frequency $f_{0}$ of tips also increases with the number $N_{arms}$ of arms. These imply that the quantities $f_{0}$, $f_{1}$, $f_{2}$, and $N_{arms}$ satisfy the following relation of Eq. (5):
\begin{eqnarray}
  f_{2}-f_{1}=N_{arms}f_{0}.
\end{eqnarray}
The applicability of Eq. (5) is explored in Fig. 8 which shows the Fourier spectra of the corresponding time sequences of 3-armed dense spiral in Fig. 7. The rotation frequency of tip is $f_{0}$$=$$0.046$ as indicated in Fig. 8(a). The frequency of spiral is $f_{1}$$=$$1.602$ as indicated in Fig. 8(b), which is far more than the rotation frequency $f_{0}$ of tip. The oscillating frequency $f_{2}$ of one point far from the spiral center is $f_{2}$$=$$1.741$ as shown in Fig. 8(c). These frequencies satisfy the relations of Eq. (5).

\indent Because the front and the back reconstruct twice during one cycle as illustrated in Fig. 6, it results in the frequency doubling $2f_{1}$ in the power spectra of distance $d$ between two neighboring tips [as indicated in Fig. 8(d)]. Meanwhile, because the tips are meandering clockwise, the frequency $f_{4}$ of zigzagged displacement of tip is the sum of the frequency doubling $2f_{1}$ and the rotation frequency $f_{0}$ of tip:

\begin{eqnarray}
  f_{4}=2f_{1}+f_{0}.
\end{eqnarray}
In Fig. 8(a), it reads $f_{4}$$=$$3.24$. In addition, the lower frequency $f_{3}$$=$$0.185$ in Fig. 8(d) is always four times the rotation frequency $f_{0}$ of tip:
\begin{eqnarray}
  f_{3}=4f_{0},
\end{eqnarray}
which should result from the boundary constraint of the square domain to the tips.

\indent These relations also hold for other multi-armed dense spirals in our extended simulations, and are not limited to the case of 3-armed dense spiral as shown in Figs. 7 and 8. Because the frequency of spiral is determined by the control parameters of the model, the frequency $f_{1}$ of spiral is almost invariant for multi-armed spirals with different numbers of arms as long as the parameter $\varepsilon$ keeps constant. The other frequencies, $f_{0}$, $f_{2}$, $f_{3}$, and $f_{4}$, increase with the number $N_{arms}$ of arms.

\indent The parameters $\varepsilon$, $a$, and $\delta$ control the Nonequilibrium Ising-Bloch bifurcation which plays important role on pattern formation in bistable media \cite{Hagberg, Hagberg1}. In the Bloch region, the speeds $c$ of the Bloch interfaces (front and back) are determined by the following implicit relation \cite{He2, Hagberg2}:

\begin{eqnarray}
  \varepsilon=\Big[\frac{1+2a}{\sqrt{8a}q^{2}(c^{2}/4+q^{2})^{1/2}}\Big]^{2}/\delta,
\end{eqnarray}
here, $q^{2}$$=$$1$$+$$\frac{1}{2a}$. It shows that decreasing $\delta$ or increasing $a$ is equivalent to increasing $\varepsilon$. In this paper, we use $\varepsilon$ as the main control parameter given that $a$$=$$2$ and $\delta$$=$$0.05$. In our extended simulations, similar interactions of multi-armed spirals have been observed when changing the parameter $a$ and $\delta$ in the parameter ranges: $1.8$$<$$a$$<$$2.5$ and $0.01$$<$$\delta$$<$$0.1$. However, the parameter $a$ could not be set to a large value in the numerical simulation because the absolute values $(|u_{\pm}|, |v_{\pm}|)$ of the two stationary and uniform stable states of system will increase with the parameter $a$. The more the absolute values $(|u_{\pm}|, |v_{\pm}|)$ are, the steeper the interfaces (front and back) connected the two stable states become, that will require numerical simulations of smaller space step and demanding computational intensity. Extended numerical simulations have also shown our analyses are applicable for general interaction mechanism between multi-armed spirals in bistable media.

\subsection{Multi-armed sparse spirals}

\indent Now, we focus on the interaction of multi-armed sparse spirals [Fig. 9], i.e. $\varepsilon$$\geq$$0.33$, which is somewhat different from the case of multi-armed dense spirals. In Fig. 3, the circle dotted line $r_{0}$ represents the radius of the tip trajectory of 1-armed spiral. When $\varepsilon$$\geq$$0.4$ the dashed lines $R$ coincide with the circle dotted line $r_{0}$. This means that the rigid rotation of an individual arm within multi-armed sparse spiral [Fig. 9] is identical to that of 1-armed sparse spiral [Figs. 1(b), 1(c), 1(f), 1(g)]. The tips travel along their individual trajectories, i.e., not always along a rough circle. This is very different from the case of multi-armed dense spirals as illustrated by Fig. 4. In that case, the tips of multi-armed dense spirals corotate and meander zigzaggedly along a rough circle. It shows that the interaction between the arms of dense multi-armed spiral is stronger than that of sparse multi-armed spiral. In addition, when $r$$<$$r_{c}$, the tips only exhibit attractive interaction [Figs. 9(a) and 9(c)]. When $r$$>$$r_{c}$, the tips will develop independently as shown by Figs. 9(b) and 9(d), which is similar to the case of dense multi-armed spiral.

\begin{figure}[htbp]
  \begin{center}\includegraphics[width=7cm,height=6cm]{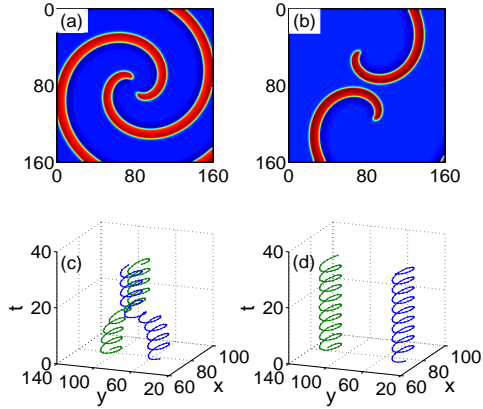}
  \caption{(Color online) Sparse 2-armed spirals and their trajectories of tips. The initial separation of the two tips is 56 space units in (a) and (c), 80 space units in (b) and (d). The parameter $\varepsilon$$=$$0.4$. (a) and (c) show the attractive process of the two tips. (b) and (d) show the independent rotation of each tip.}
  \label{9}
  \end{center}
\end{figure}

\section{CONCLUSION AND REMARKS}
\indent In this work, we have studied the interaction of multi-armed spirals in bistable media. In order to illustrate this interaction, we first give the evolution of 1-armed spiral as shown in Fig. 1. Multi-armed spirals can be obtained by constructing appropriate initial conditions. For dense multi-armed spirals, when $r$$<$$r_{c}$, strong interaction between arms results in the corotation of tips along a zigzagged circle. The tips will keep one wavelength apart as they corotate, which is very different from the case of multi-armed spiral in excitable media. During the spiral rotating, the arms collide, connect and disconnect continuously, which drives the tips to meander zigzaggedly. In one cycle, the front and the back reconstruct twice, the arms switch with the tips once. For dense multi-armed spirals, the rotation frequency $f_{0}$ of tip, the frequency $f_{1}$ of spiral, the oscillation frequency $f_{2}$ of media, the frequency $f_{4}$ of zigzagged displacement, and the number $N_{arms}$ of arms satisfy the general relations of Eqs. (5)-(7). When $r$$>$$r_{c}$, the arms in multi-armed dense spirals will rotate independently just like 1-armed dense spiral. For sparse multi-armed spirals, the rigid rotation of an individual arm within multi-armed sparse spiral is identical to that of 1-armed sparse spiral. When $r$$<$$r_{c}$, the tips exhibit only attractive interaction and could rotate along several circles, which is different from the case of dense multi-armed spirals.

\indent It should be mentioned that our numerical results are robust against little deviations of the initial conditions used, which makes it reliable to perform correlative studies experimentally. The presented results reveal the general mechanism of interaction of multi-armed spirals in bistable media, and are not limited to the specific cases illustrated by Figs. (4)-(9). Near the bifurcation point $\varepsilon_{c}$$=$$0.33$ (the patterned rectangle region in Fig. 3), the dynamics of multi-armed spirals will become very complex. In order to investigate this region, higher spatial-temporal resolution (i.e., large domain size, small space and time steps) are fairly necessary. This will need numerical simulations of demanding computational intensity.

\indent As is well known pattern formations in bistable media are related to the initial conditions besides the control parameters. In our work, we generated the multi-armed spirals by designing appropriate initial conditions and then studied their interactions. Many real chemical experiments, such as the ferroin-, Ru(bpy)$_{3}$-, and cerium-catalyzed Belousov-Zhabotinsky and/or iodate-sulfite reactions, are sensitive to visible and/or ultraviolet light \cite{Lee1,Lee2,Kheowan,Vanag,Zykov}. Therefore, light illumination has been widely used to excite desired initial conditions. For example, a couple of spirals can be generated by shadowing a small part of the propagating wave front with a mask. Multi-armed spirals with corotation and independent rotation have been observed experimentally by Steinbock \cite{Steinbock} and Krinsky \cite{Krinsky2} in excitable media. Light-sensitive catalyst even makes it possible to study the image processing and the effective computational procedure for finding paths in labyrinth based on reaction-diffusion mechanisms \cite{Kuhnert}. At present, bistable patterns, such as the 1-armed spirals, the labyrinthine, and the breathing spots, have been studied well in iodate-sulfite reactions experimentally \cite{Judit,Boissonade,Lee1,Lee2} and in FitzHugh-Nagumo model theoretically \cite{Hagberg,Hagberg2}. We believe this work provide information to understand the interaction of multi-armed spiral, and hope that our results about the interaction of multi-armed spirals in bistable media can be verified in one of these light-sensitive reactions. For example, multi-armed spiral should be observed by illuminating the ferrocyanide-iodate-sulfite medium with ultraviolet light if employing our suggested initial conditions.

\section{ACKNOWLEDGMENTS}
\indent  This work is supported by the National Natural Science Foundation of China (Grant Nos. 11205044, 11175067), the Natural Science Foundation of Hebei Province, China (Grants Nos. A2011201006, A2012201015), the Research Foundation of Education Bureau of Hebei Province, China (Grant No. Y2012009), and the Science Foundation of Hebei University (Grant No. 2011JQ04).

\end{document}